\documentstyle[12pt,aasms4]{article}

\def\etal{{\it et al. }\ }

\def\apj    {ApJ{\rm,}\ }

\def\aap    {A{\rm\&}A{\rm,}\ }

\def\mnras  {MNRAS{\rm,}\ }
\def\nat    {Nat{\rm,}\ }

\def\sci    {Sci{\rm,}\ }

\begin{document}

\title{INTRINSIC SIZE OF SGR~A$^\star$: 72 SCHWARZSCHILD RADII}

\author{K.~Y. Lo}
\affil{Academia Sinica Institute of Astronomy \& Astrophysics, PO Box 1-87, 
Nankang, Taipei~115; kyl@asiaa.sinica.edu.tw}

\author{Zhi-Qiang Shen\altaffilmark{1}}
\affil{Academia Sinica Institute of Astronomy \& Astrophysics, PO Box 1-87, 
Nankang, Taipei~115; and Shanghai Observatory, 80 Nandan Road, Shanghai 200030}

\and

\author{Jun-Hui Zhao and P.~T.~P. Ho}
\affil{Harvard-Smithsonian Center for Astrophysics, 60 Garden Street, 
Cambridge, MA 02138; jzhao@cfa.harvard.edu and ho@cfa.harvard.edu}

\altaffiltext{1}{Present address: National Astronomical Observatory, Mitaka, 
Tokyo 181; zshen@hotaka.mtk.nao.ac.jp}

\begin{abstract}

Recent proper motion studies of stars at the very center of the Galaxy
strongly suggest that Sagittarius (Sgr)~A${\bf 
^\star}$, the compact nonthermal radio source at the Galactic Center,  is 
a 2.5${\times 10^6~{\rm M_\odot}}$ black hole.
By means of near-simultaneous multi-wavelength Very Long Baseline Array 
measurements, we determine for the first time 
the intrinsic size  and shape of Sgr~A${\bf ^\star}$ to be 
72~R$_{\bf\rm sc}(\star)$ by $<$ 20~R$_{\bf\rm sc}(\star)$, with the
major axis oriented essentially north-south,  
where R$_{\bf\rm sc}(\star)~(\equiv ~ 7.5 \times 10^{11}$~cm)
is the Schwarzschild radius for a 2.5${\times 10^6~{\rm M_\odot}}$ black hole. 
Contrary to previous expectation that the intrinsic structure of 
Sgr~A${\bf ^\star}$ is observable only at $\lambda 
\le$~1~mm, we can discern the intrinsic source size at $\lambda$ 7~mm
because 
(1) the scattering size along the minor axis is half that along the major axis, 
and (2)
the near simultaneous multi-wavelength mapping of Sgr~A$^\star$
with the same interferometer makes it 
possible to 
extrapolate precisely the minor axis scattering angle at $\lambda$ 7~mm.
The 
intrinsic
size and shape place direct constraints on the various emission models for 
Sgr~A${\bf ^\star}$.
In particular, the advection dominated accretion flow model may have to 
incorporate a radio jet 
in order to account for the structure of Sgr~A${\bf ^\star}$.

\end{abstract}


\section{Introduction}

Sagittarius (Sgr)~A${\bf ^\star}$, the extremely compact 
non-thermal radio source at the Galactic Center, has been for many years 
considered the signpost of a massive black hole (e.g. Lynden-Bell and Rees 1971; 
Lo \etal 1985). Recent proper motion 
and radial velocity measurements of the stars in the immediate neighborhood 
of Sgr~A${\bf ^\star}$ have provided very compelling dynamical evidence for 
the existence of a compact dark mass of 2.5${\times 10^6~{\rm M_\odot}}$ 
located within 0.015~pc (4.5${\times 10^{16}}$~cm) of Sgr~A${\bf ^\star}$, 
supporting the hypothesis that Sgr~A${\bf ^\star}$ is powered by a single 
massive black hole (Eckart and Genzel 1996; Eckart and Genzel 1997; Ghez \etal 
1998).  Since Sgr~A${\bf ^\star}$ would be the nearest example by far of such
a system, determining its intrinsic source structure 
would be very important for probing the region immediately 
surrounding the massive black hole.  

Up till now, Very Long Baseline Interferometric (VLBI) observations of 
Sgr~A${\bf ^\star}$ have not been able to probe its intrinsic structure,
due to the scattering by the interstellar electrons (e.g. Davies et al, 1976; 
Lo et al 1981, 1985, 1993; Backer \etal 1993; Rogers \etal 1994; 
Krichbaum \etal 1997; Bower and Backer 1998).  In this paper, 
we report our efforts to image
Sgr~A${\bf ^\star}$ with the Very Long Baseline Array (VLBA) nearly 
simultaneously at five wavelengths (${\bf \lambda = }$6.0, 3.6, 2.0, 1.35~cm and 
7~mm).  
The {\sl multi-wavelength} imaging,
with the same interferometer, 
is crucial for our differentiating
interstellar scattering effects from the intrinsic source structure of Sgr A*. 
Finally, more than 20 years after its discovery (Balick and Brown 1974), 
we have for the first time determined
that  the intrinsic size of 
Sgr~A${\bf ^\star}$ is asymmetric,
being 72~R$_{\bf\rm sc}(\star)$ by $<$ 20~R$_{\bf\rm 
sc}(\star)$,
where R$_{\bf\rm sc}(\star)~(\equiv ~ 7.5 \times 10^{11}$~cm)
is the Schwarzschild radius of a 2.5${\times 10^6~{\rm M}_\odot}$~black hole.  
The new VLBA result is consistent with the latest millimeter wavelength
VLBI results at 3.5 and 1.4 mm, with 3 and 1 baseline respectively,
which suggests that the intrinsic size of Sgr A* may be 
17$\pm$9 R$_{\bf\rm sc}(\star)$ (Krichbaum, T. P. etal 1998).

\section{Observations and Data Analysis}

The observations were carried out 
using eleven National Radio Astronomy
Observatory (NRAO)\footnote {The NRAO is operated by Associated Universities
Inc., under cooperative agreement with the National Science Foundation} 
25-m radio telescopes: 10 VLBA antennas and 1 VLA antenna.
The 7~mm observations were carried out on 14 February 1997,
while the 1.35~cm observations were interlaced with the 2.0~cm observations
on 12 February 1997, and the 3.6~cm interlaced with the 6.0~cm on 7 February 
1997,
all at the same hour ranges of UT12h00-20h00.
Quasars, NRAO~530 and PKS~1921-293, served as amplitude calibrators and fringe
detection sources.  The data recording was in the standard VLBA mode,
with 32 MHz bandwidth for both circular polarizations at each telescope site. 
Correlation of the data was done with the VLBA correlator in Socorro, 
New Mexico.
All the post-correlation reduction was carried out by using the NRAO AIPS
and the Caltech VLBI package including Difmap.  With global fringe fitting, 
Sgr~A${\bf ^\star}$ was detected on the short and intermediate baselines 
depending
on the observed wavelength, whereas both calibrators (NRAO~530 and 
PKS~1921-293)
were detected on all the baselines.   
The visibility amplitude calibrations were done using system temperature 
measurements at each site.  At $\lambda$7~mm, special attention was paid
to the elevation dependent opacity corrections, while at $\lambda$$\ge$1.35~cm 
the atmospheric opacity was not significant.   Images of Sgr~A${\bf ^\star}$
were produced for all five wavelengths using standard hybrid mapping.  

An elliptical Gaussian model was fitted by the least-squares method to both 
amplitudes and phases in the calibrated visibility data to yield a quantitative 
description of the source structure.
Fig.~1 illustrates the model fitting to the 7 mm data, showing both
the amplitude as a function of visibilities and the closure phase triangles.
We emphasize by this figure the good quality of the data, the availability
of many baseline where the structure can be fitted, and the robustness
of the fit. The steadily improved performance of the VLBA played a pivotal
role in the quality of this data set.
The total flux density (S$_{\nu}$), FWHM major axis diameter ($\theta_{\rm 
major}$), 
FWHM minor axis diameter ($\theta_{\rm minor}$), its axial ratio of minor to 
major 
axis diameters, and the position 
angle (P.A.) of the major axis of the model fit are given in Table~1, in which 
we
also include previous 7 mm results from 1994.74 for comparison (Bower and Backer 
1998).  
At wavelengths $\ge$1.35~cm, the mean P.A. of the major axis is $80^\circ 
\pm 3^\circ$ (essentially E-W) and the mean axial ratio is 0.53 $\pm$ 0.07.  
At 7~mm, the axial ratio, 0.83$\pm$0.11, is significantly different from that 
at the longer wavelengths, while the P.A. is not. 

\section{The Intrinsic Structure of Sgr~A$^\star$}

Our near-simultaneous multi-wavelength VLBA mapping of Sgr~A$^\star$ allows 
us to plot {\sl both} the measured major and minor axis diameters versus the 
observing wavelength in Fig.~2. 
The measured major axis diameters (open circles) can be fit by 
$\theta_{\rm major} =(1.43 \pm 0.02)~\lambda^{1.99\pm0.03}$~marc s ($\lambda$ in 
cm), 
represented by the solid line.  
Within the accuracy of the experiment, such an index is indistinguishable from 
2, 
so that the major axis diameters appear to follow a $\lambda^{2}$ law 
over a range of $\lambda$ from 7~mm to 6~cm, in excellent agreement with the 
previous result of 1.42~$\lambda^{2.0}$ (e.g. Alberdi \etal 1993). 
This $\lambda^2$-dependence is 
consistent with the measured size being dominated 
by the scattering angle that is a result of the radiation from Sgr~A$^\star$ 
propagating through the interstellar medium with fluctuations 
in the electron density. The 
power spectrum of the density fluctuation of interstellar electrons is 
normally assumed to be $\propto {\rm k}^{-\beta}$, where k is the wavenumber of 
the irregularities.  The scattering angle scales as
$\lambda^{1 + 2/(\beta - 2)}$, where $(1 + 2/(\beta - 2)) =$ 2, 2.2 for
$\beta = 4, 11/3$ respectively (Romani, Narayan, and Blandford 1986).

Along the minor axis, a fit to the measurements at all five wavelengths 
yields $\theta_{\rm minor} = (1.06\pm0.10)~\lambda^{1.76\pm0.07}$~marc s
which appears inconsistent with interstellar scattering.  
A fit to all points for $\lambda \ge$1.35~cm,
however, yields a dependence of
(0.87$\pm$0.23)~$\lambda^{1.87\pm0.16}$~marc s, which is consistent, 
within the errors, with
the $\lambda^2$-dependence expected for interstellar scattering.
If we assume $\lambda^2$-dependence derived for the major axis also applies to 
the minor axis, 
$\theta_{\rm minor} = (0.76\pm0.05)~\lambda^{2.0}$ for $\lambda \ge$~1.35~cm (cf 
Fig.~1).  
This dependence also agrees very well with 
$\theta_{\rm major} = (1.43~\pm 0.02)\lambda^{1.99\pm 0.03}$ and a constant 
axial 
ratio of 0.53.

{\sl This is the first time the $\lambda$-dependence of the minor axis diameters 
is determined directly by observations}, the results of which strongly suggest 
interstellar scattering dominates the observed minor axis image size at $\lambda 
\ge$1.35~cm.
The elongation of the scatter-broadened image can be caused by an anisotropic 
scattering medium in the vicinity of the Galactic center.  The anisotropy
in the electron fluctuations in the interstellar medium has been postulated to 
be
due to turbulence in a magnetized plasma, and the mechanism to generate the 
density
fluctuation has been proposed to be due to the specific entropy being mixed by 
shear Alfvenic
turbulence that has "eddies" elongated in the direction of the magnetic field on 
small
spatial scales (Higdon 1984; Goldreich and Sridhar 1995).

Importantly, Fig.~2 shows a significant deviation at 7~mm 
between the measured minor axis diameter, $\theta _{\rm minor}$, 
and the scattering angle, $\theta_{\rm sc}$, extrapolated from 
0.76~$\lambda^{2.0}$: $\Delta\theta \equiv 
\theta _{\rm minor} - \theta_{\rm sc} 
= (0.58~\pm~0.07)$~marc s $- (0.37~\pm ~0.02)$~marc s $= (0.21~\pm~0.07)$~marc 
s.
The deviation of $\theta _{\rm obs}$
from the $\lambda^2$ dependence is naturally expected when the intrinsic source 
diameter, $\theta_{\rm int}$, becomes comparable to the scattering angle, 
since $\theta _{\rm minor} = \sqrt{\theta _{\rm int}^2 + \theta_{\rm sc}^2}$
(Narayan and Hubbard 1988).
{\sl Thus, this implies that at 7 mm, $\theta_{\rm int}$ = 
(0.45~$\pm$~0.11)~marc s for 
Sgr~A$^\star$ along the minor axis (P.A. = $-10^\circ$; nearly N-S) direction.} 
If we use the measurements of Bower and Backer (1998), 
$\theta _{\rm minor} = (0.55~\pm~0.11)$~marc s, $\Delta\theta = 
(0.18~\pm~0.11)$~marc s 
~and $\theta_{\rm int} = (0.41~\pm~0.17)$~marc s.  Combining the two sets of 
measurements, 
we obtain $\theta _{\rm minor} = (0.57~\pm~0.06)$~marc s, $\Delta\theta = 
(0.20~\pm~0.06)$~marc s
and $\theta_{\rm int} = (0.44~\pm~0.09)$~marc~s. We note that this intrinsic 
size scale
is larger than the value inferred at 1.4 mm by Krichbaum etal (1998).
In addition to a possible $\lambda$ dependence on the intrinsic
size, we note that the 1.4 mm measurements were based on
a single baseline.

The reasons that we can discern the intrinsic source size at $\lambda$ 7~mm, 
contrary to previous 
expectations that intrinsic source size is observable only at $\lambda 
\le$~1~mm, are 
(1) the scattering size along the minor axis is half that along the major axis, 
and (2)
the near simultaneous multi-wavelength mapping of Sgr~A$^\star$,
with the same instrument, over the same hour angle,
and calibrated in a uniform manner,  makes it 
possible to 
extrapolate precisely the minor axis scattering angle at $\lambda$ 7~mm.

Conceivably, changes in the refractive properties of the interstellar medium 
could lead 
to the deviation indicated above, since the 7~mm refractive scattering time 
scale for 
Sgr~A$^\star$ (proportional to  $\lambda^2$) could be short:   
t$_{\rm ref} = \theta_{\rm sc}{\rm D/V \le 0.5 \times (10~kms^{-1}/V)}$ year, 
where 
D, and V are the distance to the scattering medium, 
the relative velocity of the scattering medium and the observer, respectively. 
However, since the 7~mm source parameters did not change over 2.4 years and 
probably 
over a longer period, the deviation of the minor axis size from 
0.76~$\lambda^{2.0}$ is 
unlikely to be due to changing refractive scattering effects at 7~mm.  
Furthermore, other evidence suggests that the refractive scattering effects for 
the Galactic 
Center must be very small (Romani, Narayan, and Blandford 1986).

Along the major axis (P.A. of 80$^\circ$; essentially E-W) direction, the 
measured diameter 
of $0.7\pm0.01$ marc s and the extrapolated scattering size of 
$0.69\pm0.01$~marc s imply that the 
intrinsic size along the same direction has to be $\le 0.13$ marc s.  
Combined 
with the 
minor axis intrinsic diameter
of 0.44 $\pm$ 0.09 marc s derived above, this implies that 
{\sl the intrinsic source structure of Sgr~A$^\star$ could be elongated along an 
essentially N-S direction, with an axial ratio of $< 0.3$.}  This also implies 
an intrinsic 
brightness temperature of $> 1.3 \times 10^{10}$ K.

Up till now, there exist only limits to the intrinsic size of Sgr~A$^\star$.
Given that the intrinsic source size of Sgr~A$^\star$ can now be estimated at 
7mm,
we can also ask whether there are constraints on the wavelength dependence
of the intrinsic source size.  
At $\lambda$3.5~mm, the upper limit to the observed size is 0.2 ~marc s, 
from which we can infer an upper limit to the intrinsic size 
along the minor axis of $< ~0.18$~marc s (Rogers \etal 1994).
At $\lambda$1.4~mm, the marginal detection  
with an interferometer with a fringe spacing of $\sim$0.3~marc s suggests a 
size scale of 0.05-0.15~marc s (Krichbaum \etal 1997; 1998).  From the absence of 
refractive scintillation 
due to focusing and defocusing of the scattered image by large scale plasma 
fluctuations (Gwinn \etal 1991)
at $\lambda$1.3~mm and 0.8~mm, the respective lower limits to the 
intrinsic size are 0.02~marc s and 0.008~marc s.  Taken all together, a 
$\lambda^{\alpha}$-dependence of $\theta_{\rm int}$, with $1.9 > \alpha > 0.7$ 
is
not inconsistent with the above limits, at least along the minor axis direction. 
Clearly, 
this very preliminary determination of the wavelength dependence of the 
intrinsic source
size has to be improved with further observations.

\section{The Emission Mechanism for Sgr~A$^\star$}

There have been several models for the structure and mechanism of radio emission 
from 
Sgr~A$^\star$. They typically involve synchrotron emission: from a pulsar wind 
that is confined 
by ram pressure (Reynolds and McKee 1980), from thermal electrons heated by the 
dissipation of magnetic 
energy as the mass-loss in the winds from stars in the vicinity of Sgr~A$^\star$ 
such as IRS16 
is spherically accreted by a massive black hole (Melia, Jokipii, and Narayanan 
1992;
Melia 1994), from a jet in a coupled 
jet-disk system (Falcke, Mannheim, and Biermann 1993), or from the 
thermal electrons at an electron temperature of $\sim 10^{9.5}$ K 
of a two-temperature plasma in a rotating advection-dominated accretion flow
(Narayan \etal 1998).

Until now, in the absence of the intrinsic source structure, for confirmation
the various models rely on comparison to the 
spectral energy distribution (SED) of Sgr~A$^\star$ from the radio wavelength to 
the $\gamma$-ray
(Melia 1994; Narayan \etal 1998; Duschl and Lesch 1994; Serabyn \etal 1997; 
Falcke \etal 1998).
However, because of the insufficient angular resolution in the wavelength bands 
shortward of $\lambda$1~mm, it is uncertain that all the radiation in the SED 
actually originates from 
Sgr~A$^\star$, making the comparison less than definitive. In contrast, the 
intrinsic 
source structure derived here provides direct spatial constraints on the various 
models.
Specifically, the spherical accretion model (Melia 1994) 
predicts a $\lambda$~7~mm size too large to be 
consistent with the results here, while the coupled disk-jet model (Falcke, 
Mannheim and
Biermann 1993) predicts a 
jet size scale smaller than the 72${\rm R_{sc}(\star)}$ obtained here.
The advection dominated accretion flow (ADAF) model for Sgr~A$^\star$
can naturally account for the luminosity far below that implied by the estimated 
accretion 
rate (Narayan \etal 1998).
However, if the $\lambda$~7 mm radiation from Sgr~A$^\star$ originates as 
thermal 
synchrotron radiation 
from the inner part of the advection dominated accretion disk, the current model 
has 
difficulty explaining the elongated shape with an axial ratio $< 0.3$ and the 
brightness 
temperature of
$> 1.3 \times 10^{10}$ K.  An additional component, a radio jet, may be needed 
in the ADAF model
to account for the intrinsic elongation of Sgr~A$^\star$.  Obviously, 
observations of the intrinsic 
source structure at $\lambda <$~7~mm will probe ever closer to the event horizon 
of the 
massive black hole
and will provide further constraints and stimuli for the models of 
Sgr~A$^\star$.  

Given the proximity of the center of our Galaxy, understanding the radio 
emission from Sgr~A$^\star$ 
provides a unique opportunity for probing the physical conditions to within $<20 
{\rm R_{sc}(\star)}$ 
of a 2.5$\times 10^6~{\rm M_\odot}$ black hole.  

\acknowledgments

We thank R. Narayan and T. Chiueh for helpful discussions. Z. Shen thanks the
Su-Shu Huang Astrophysics Research Foundation for travel support to the Array 
Operations Center for data reduction. Research at the ASIAA is funded by the 
Academia Sinica in Taipei.

\clearpage


\figcaption{A plot shows the Gaussian elliptical 
model fitting (solid curves) to the visibility
data  at 7 mm (vertical bars). Left panels: amplitude vs. baseline pairs
(BR-FD, BR-KP, FD-OV, FD-KP, OV-Y, and  FD-LA).
The maximum baseline lengths of the pairs are 2346, 1914, 1508, 744, 1025
and 607 kilometers, respectively. BR-FD is  the longest baseline in 
NS. Right panels: closure phase triangles. \label{fig1}}

\figcaption{A log-log plot of measured (FWHM) source size versus observing 
wavelength for Sgr~A$^\star$ (7- 14 February 1997). The solid line represents 
a 1.43~$\lambda$$^{1.99}$ fit to the major axis sizes (open circles), while 
the dashed line a 0.76~$\lambda$$^{2.0}$ fit 
to the minor axis sizes (filled circles). \label{fig2}}

\clearpage

\medskip
\medskip
\medskip
\medskip
\begin{center}
{\bf Table 1.  Parameters of Elliptical Gaussian Model Fit}
\medskip

\begin{tabular}{ccccccc} 
~~\\
~~\\

\hline
 $\lambda$ &  $\nu$   & S$_\nu$ & $\theta_{\rm major}$ & $\theta_{\rm minor}$ & 
Axial &   P.A. \\ 
     (cm)  &  (GHz)   &   (Jy)  &     (marc s)     &     (marc s)     & Ratio & 
($^\circ$) \\ 
\hline
6.03 & 4.97 & 0.60$\pm$0.09 & 49.6$\pm$4.50 & 25.1$\pm$2.00 & 0.51$\pm$0.09 & 
81$\pm$3     \\
3.56 & 8.41 & 0.73$\pm$0.10 & 18.0$\pm$1.53 & 9.88$\pm$1.68 & 0.55$\pm$0.14 & 
78$\pm$6     \\
1.96 & 15.3 & 0.68$\pm$0.06 & 5.84$\pm$0.48 & 3.13$\pm$1.14 & 0.54$\pm$0.21 & 
73$\pm$14    \\
1.35 & 22.2 & 0.74$\pm$0.04 & 2.70$\pm$0.15 & 1.50$\pm$0.59 & 0.56$\pm$0.25 & 
81$\pm$11    \\
0.69 & 43.2 & 1.03$\pm$0.01 & 0.70$\pm$0.01 & 0.58$\pm$0.07 & 0.83$\pm$0.11 & 
87$\pm$8     \\
0.69$^{\dag}$ & 43.2 & 1.28$\pm$0.1 & 0.76$\pm$0.04 & 0.55$\pm$0.11 & 0.73$\pm$0.1 
& 77$\pm$7     \\

\hline
\end{tabular} 
\end{center}
\medskip
\medskip
{\small ~$^\dag$ For comparison, the corresponding results at 43.2 GHz from 
1994.74 (Bower and Backer 1998).}


\begin{thebibliography}{}
\bibitem{}{Alberdi, A. \etal 1993, \aap 277, L1}
\bibitem{}{Backer, D. C., Zensus, J. A., Kellermann, K. I., Reid, M., 
Moran, J. M., and Lo, K. Y. 1993, \sci 262, 1414}
\bibitem{}{Balick, B., and Brown, R. 1974, \apj 194, 265}
\bibitem{}{Bower, G. C., and Backer, D. C. 1998, \apj 496, L97}
\bibitem{}{Duschl, W. J., and Lesch, H. 1994, \aap 286, 431}
\bibitem{}{Eckart, A., and Genzel, R. 1996, \nat 383, 415}
\bibitem{}{Eckart, A., and Genzel, R. 1997, \mnras 284, 576}
\bibitem{}{Falcke, H., Goss, W. M., Matsuo, H., Teuben, P., Zhao, J-H, and 
Zylka, R. 1998, \apj 499, 731}
\bibitem{}{Falcke, H., Mannheim, K., and Biermann, P. 1993, \aap 278, L1}
\bibitem{}{Ghez, A. M., Klein, B. L., Morris, M., and Becklin, E. E. 1998, 
ApJ, in press}
\bibitem{}{Goldreich, P., and Sridhar, S. 1995, \apj 438, 763}
\bibitem{}{Gwinn, C. R. \etal 1991, \apj 381, L43}
\bibitem{}{Higdon, J. C. 1984, \apj 285, 109}
\bibitem{}{Krichbaum, T. P. \etal 1997 \aap 323, L17}
\bibitem{}{Krichbaum, T. P. \etal 1998 \aap 335, L106}
\bibitem{}{Lo, K. Y. \etal 1981, \apj 249, 504}
\bibitem{}{Lo, K. Y. \etal 1985, \nat 315, 124}
\bibitem{}{Lo, K. Y. \etal 1993, \nat 362, 38}
\bibitem{}{Lynden-Bell, D., and  Rees, M. J. 1971, \mnras 152, 461}
\bibitem{}{Melia, F. 1994, \apj 426, 577}
\bibitem{}{Melia, F., Jokipii, J. R., and Narayanan, A. 1992, \apj 395, L87}
\bibitem{}{Narayan, R., and Hubbard, W. B. 1988, \apj 325, 503}
\bibitem{}{Narayan, R., Mahadevan, R., Grindlay, J., Popham, R., and 
Gammie, C. 1998, \apj 492, 554}
\bibitem{}{Reynolds, S. P.,  and McKee, C. F. 1980, \apj 239, 893}
\bibitem{}{Rogers, A. E. E. \etal 1994, \apj 434, L59}
\bibitem{}{Romani, R., Narayan, R., and Blandford, R. 1986, \mnras 220, 19}
\bibitem{}{Serabyn, E. \etal 1997, \apj 490, L77}
\end{thebibliography}
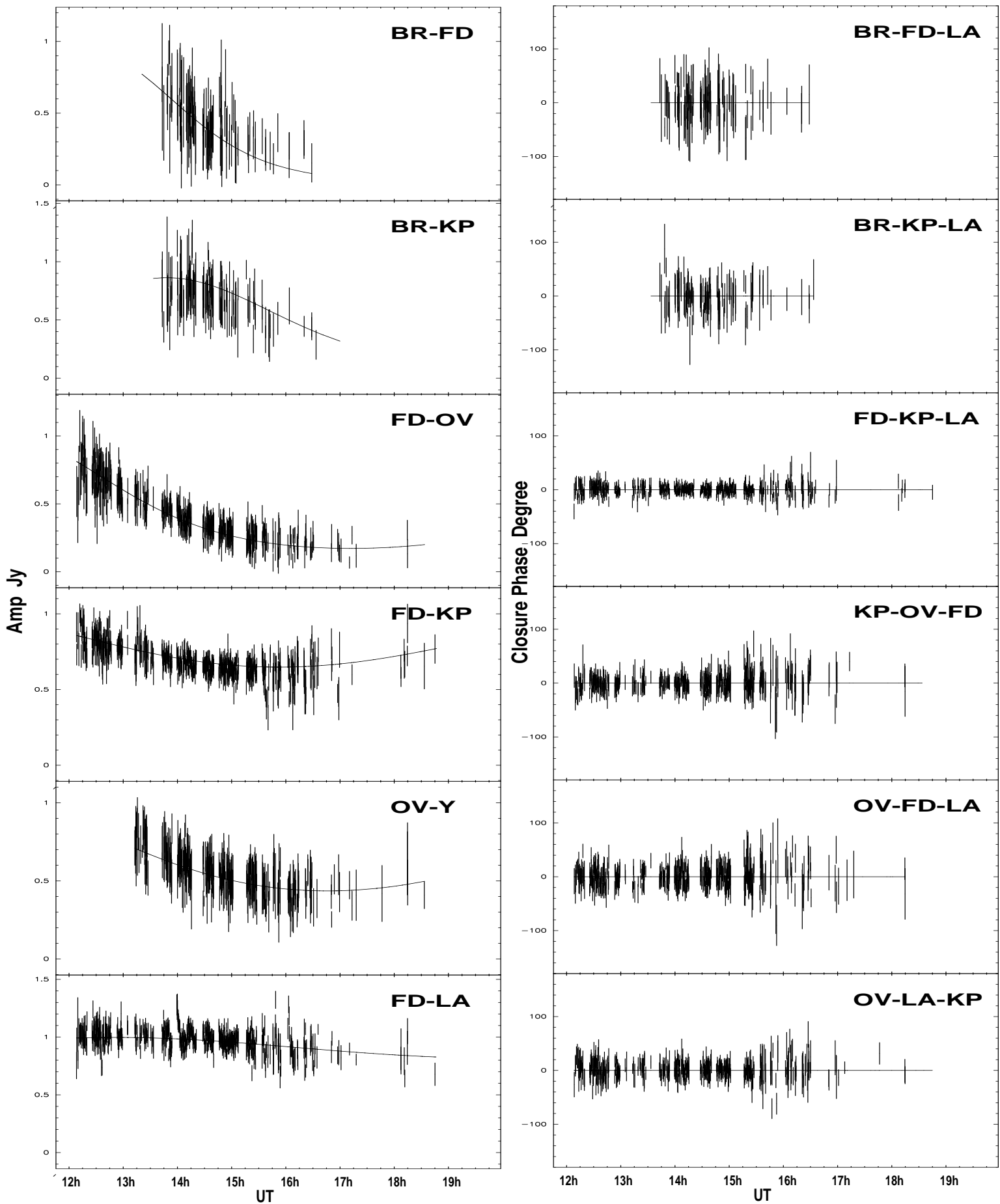
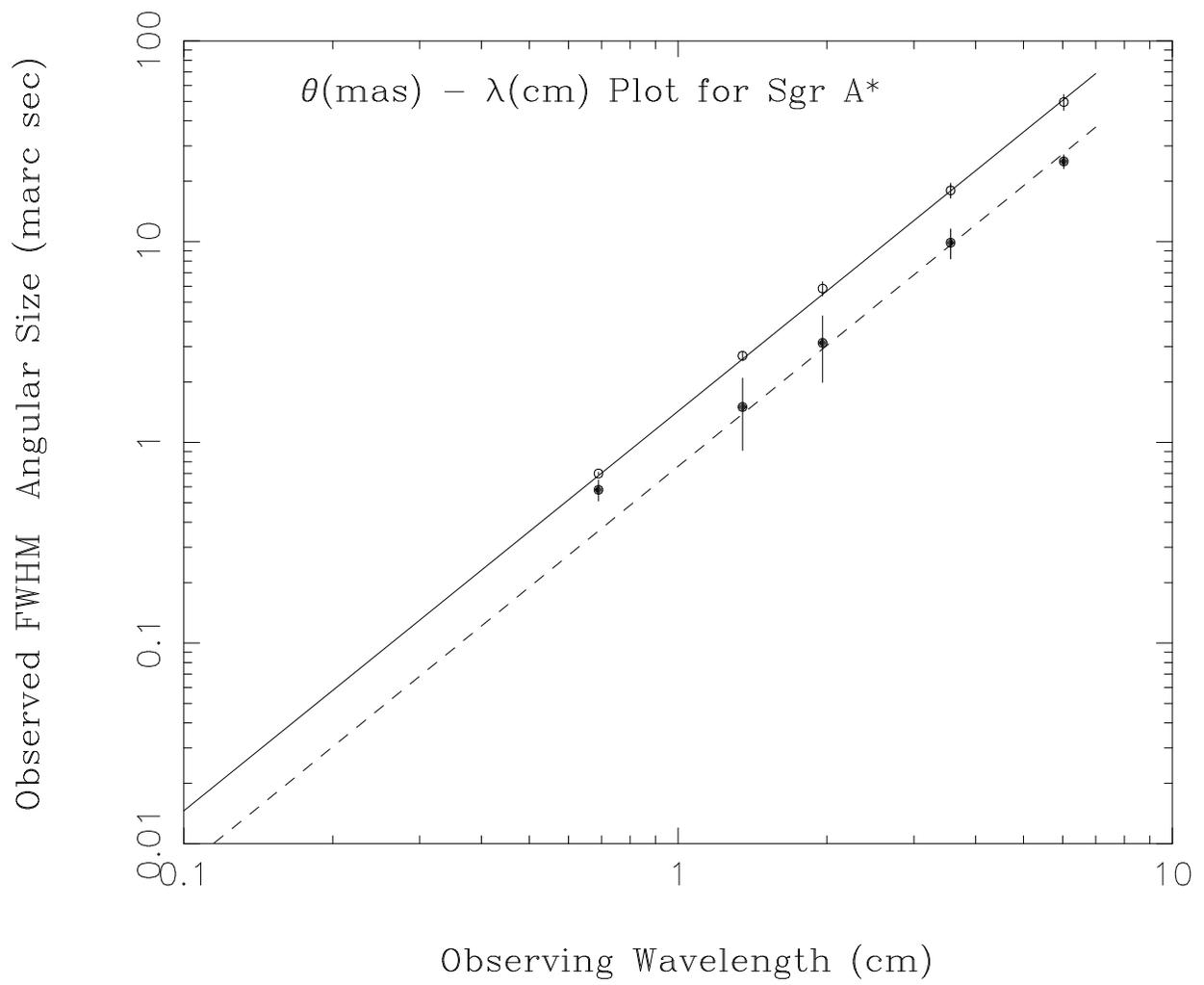
\end{document}